\documentclass[twocolumn, switch]{article} % Method A for two-column formatting

\usepackage{preprint}

\usepackage{amsmath, amsthm, amssymb, amsfonts}

\usepackage[numbers,square]{natbib}
\usepackage{siunitx}
\usepackage{tikz}%test123
\newcommand*\circled[1]{\tikz[baseline=(char.base)]{
            \node[shape=circle,draw,inner sep=2pt] (char) {#1};}}
\def\BibTeX{{\rm B\kern-.05em{\sc i\kern-.025em b}\kern-.08em
    T\kern-.1667em\lower.7ex\hbox{E}\kern-.125emX}}
\usepackage[utf8]{inputenc}	% allow utf-8 input
\usepackage[T1]{fontenc}	% use 8-bit T1 fonts
\usepackage{xcolor}		% colors for hyperlinks
\usepackage[colorlinks = true,
            linkcolor = purple,
            urlcolor  = blue,
            citecolor = cyan,
            anchorcolor = black]{hyperref}	% Color links to references, figures, etc.
\usepackage{booktabs} 		% professional-quality tables
\usepackage{nicefrac}		% compact symbols for 1/2, etc.
\usepackage{microtype}		% microtypography
\usepackage{lineno}		% Line numbers
\usepackage{float}			% Allows for figures within multicol

\usepackage{lipsum}		%  Filler text

\usepackage{newfloat}
\DeclareFloatingEnvironment[name={Supplementary Figure}]{suppfigure}
\usepackage{sidecap}
\sidecaptionvpos{figure}{c}

\usepackage{titlesec}
\titlespacing\section{0pt}{12pt plus 3pt minus 3pt}{1pt plus 1pt minus 1pt}
\titlespacing\subsection{0pt}{10pt plus 3pt minus 3pt}{1pt plus 1pt minus 1pt}
\titlespacing\subsubsection{0pt}{8pt plus 3pt minus 3pt}{1pt plus 1pt minus 1pt}

\usepackage{tikz,xcolor,hyperref}

\definecolor{lime}{HTML}{A6CE39}
\DeclareRobustCommand{\orcidicon}{
	\begin{tikzpicture}
	\draw[lime, fill=lime] (0,0) 
	circle [radius=0.16] 
	node[white] {{\fontfamily{qag}\selectfont \tiny ID}};
	\draw[white, fill=white] (-0.0625,0.095) 
	circle [radius=0.007];
	\end{tikzpicture}
	\hspace{-2mm}
}
\foreach \x in {A, ..., Z}{\expandafter\xdef\csname orcid\x\endcsname{\noexpand\href{https://orcid.org/\csname orcidauthor\x\endcsname}
			{\noexpand\orcidicon}}
}
\title{Analog Feedback-Controlled Memristor programming Circuit for analog Content Addressable Memory}

\usepackage{authblk}

\author[1,3,4]{Jiaao Yu}
\author[1,2]{Paul-Philipp Manea}
\author[1,2]{Sara Ameli}
\author[1,2]{Mohammad Hizzani}
\author[3]{Amro Eldebiky}
\author[1,2]{John Paul Strachan}

\affil[1]{PGI-14, Forschungszentrum Jülich, Aachen, Germany}
\affil[2]{RWTH Aachen University, Aachen, Germany}
\affil[3]{Technical University of Munich, Munich, Germany}
\affil[4]{Nanyang Technological University, Singapore}

\begin{document}

\twocolumn[ % Method A for two-column formatting
  \begin{@twocolumnfalse} % Method A for two-column formatting
  
\maketitle

\begin{abstract}
Recent breakthroughs in associative memories suggest that silicon memories are coming closer to human memories, especially for memristive Content Addressable Memories (CAMs) which are capable to read and write in analog values. However, the Program-Verify algorithm, the state-of-the-art memristor programming algorithm, requires frequent switching between verifying and programming memristor conductance, which brings many defects such as high dynamic power and long programming time. Here, we propose an analog feedback-controlled memristor programming circuit that makes use of a novel look-up table-based (LUT-based) programming algorithm. With the proposed algorithm, the programming and the verification of a memristor can be performed in a single-direction sequential process. Besides, we also integrated a single proposed programming circuit with eight analog CAM (aCAM) cells to build an aCAM array. We present SPICE simulations on TSMC 28nm process. The theoretical analysis shows that 1. A memristor conductance within an aCAM cell can be converted to an output boundary voltage in aCAM searching operations and 2. An output boundary voltage in aCAM searching operations can be converted to a programming data line voltage in aCAM programming operations. The simulation results of the proposed programming circuit prove the theoretical analysis and thus verify the feasibility to program memristors without frequently switching between verifying and programming the conductance. Besides, the simulation results of the proposed aCAM array show that the proposed programming circuit can be integrated into a large array architecture.
\end{abstract}
%\keywords{First keyword \and Second keyword \and More} % (optional)
\vspace{0.35cm}

  \end{@twocolumnfalse} % Method A for two-column formatting
] % Method A for two-column formatting

%\begin{multicols}{2} % Method B for two-column formatting (doesn't play well with line numbers), comment out if using method A

%%%%%%%%%%%%%%%  Main text   %%%%%%%%%%%%%%%
% \linenumbers

\section{Introduction}
Associative memories are inherent in human brains. Through association, the features of a new object can be matched with another object with similar features existing in our memory, and thus this new object can be simply remembered. In general, associative memories allow us to link uncorrelated information and hence are an essential feature of our intelligence\cite{b1}.
\par
Inspired by the biological and psychological concept of associative memories, different hardware approaches to such systems have been realized recently. W. A. Borders et al. proposed an analog spin-orbit torque device for ANN-based associative memory operation \cite{b2}. Pavlov associative memory has been implemented in circuits by different groups \cite{b3}\cite{b4}. In recent years, memristive CAMs have been regarded as critical hardware in a wide range of associative memory applications: Li et al. designed the first aCAM which allows for searching in analog values\cite{b5}; G. Pedretti et al. used aCAM to introduce a new compute concept for tree-based learning techniques (e.g., decision trees and random forests)\cite{b6}; C. E. Graves et al. proposed an 86$\times$12 memristor ternary CAM (TCAM) array for pattern matching\cite{b7}; Furthermore, differential CAM (dCAM) is also a viable option for associative memory architectures\cite{b8}. 
\par
However, good memristor programming techniques are necessary for both aCAM and dCAM arrays to achieve an accurate and fast writing process without losing high memory density. Here we present a novel LUT-based programming algorithm for memristors that utilizes an analog feedback control mechanism to enable highly precise programming of a 10T2M aCAM cell\cite{b9}. We will also show the possibility to integrate the proposed programming circuit in a 10T2M aCAM array. With the proposed programming circuit, the aCAM array can properly set, reset, or read the conductance of the memristor within a single cell, and hence provide individual addressing. Besides, our results prove that the memristor programming algorithm is possible to replace the traditional Program-Verify algorithm for memristor programming in analog CAMs.

\section{analog feedback-controlled memristor programming circuit}
\subsection{10T2M aCAM Cell}
\begin{figure}[h]
\centerline{\includegraphics[scale=0.19]{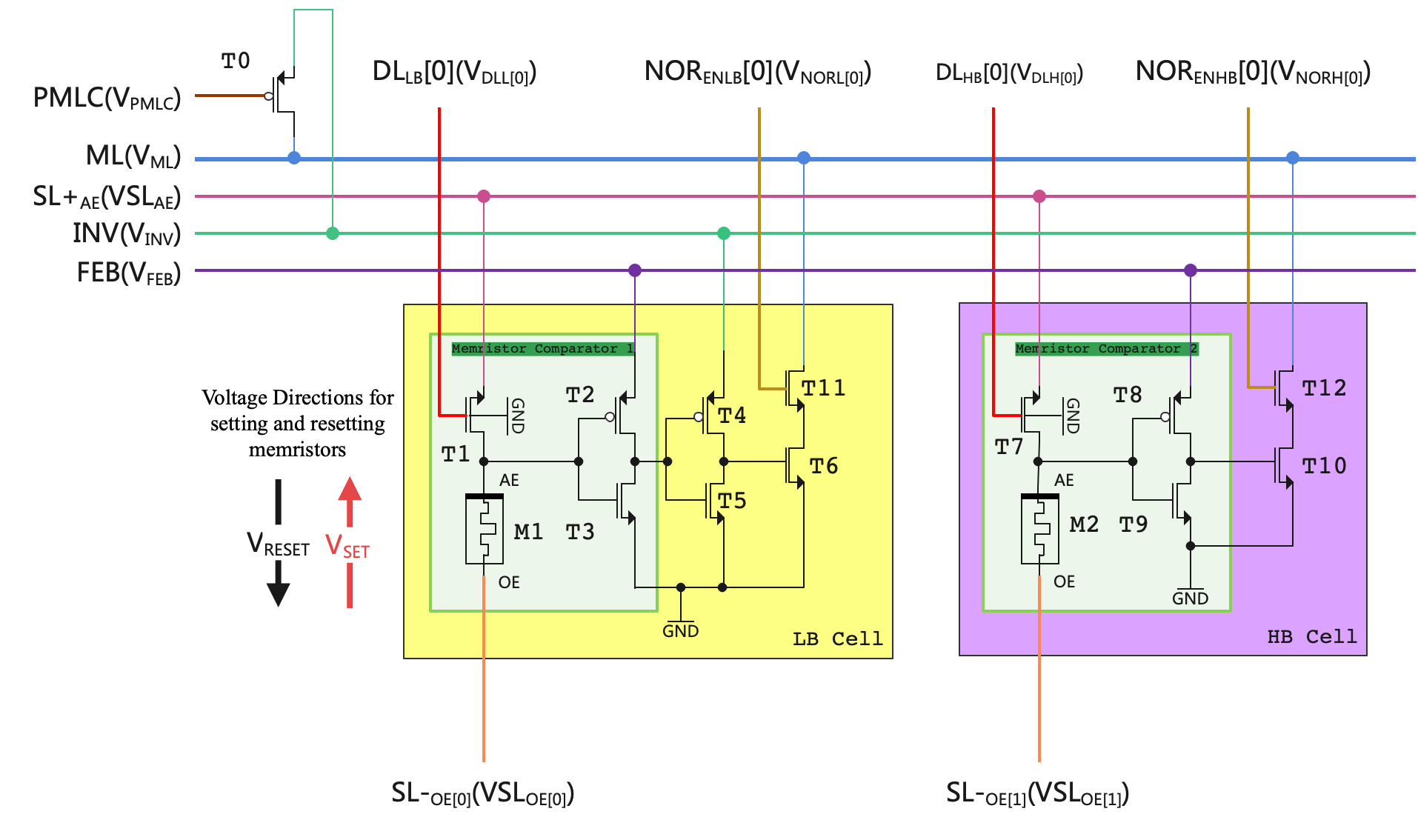}}
\caption{The 10T2M aCAM cell. This architecture consists of a low-bound (LB) cell (yellow block) and a high-bound (HB) cell (purple block) that define the boundaries of analog matching ranges. The memory cell can work in the search mode and in the program mode. In search mode, the memory performs a parallel operation to find if there is a matching word stored inside the array. In program mode, the memristor conductance is programmed to change the analog boundary of the matching word. To set the memristor, a positive voltage that exceeds the device's threshold voltage from OE to AE must be applied, while resetting requires a negative voltage.}
\label{fig1}
\end{figure}
To enhance the programming capability of the 10T2M cell proposed in \cite{b6} and \cite{b9} using our proposed algorithm, we implemented a series of modifications as shown in Fig.~\ref{fig1}, including:
\begin{enumerate}
    \item The original aCAM design utilizes 1T1R structures with memristors connected to the OE electrode, and the CAM search is performed in the reset direction. Our design features an interchanged position between the memristor and transistor T1, with the Memristor AE now connected to the transistor. This configuration results in improved performance during set operations by eliminating the body effect, when applying a set voltage. \cite{b10}
    \item The $SL_{HI}$ originally shared by all PMOS and memristors in \cite{b6}\cite{b9} are divided into 3 different signal lines ($SL+_{AE}$, INV, and FEB). This division will help the proposed memristor programming circuit to achieve the LUT-based programming algorithm and better control the analog feedback loop.
    \item T11 and T12 are added in the LB cell and the HB cell respectively. These transistors add a "don't care" option in CAM search to each cell. They also ensure in during a search operation that the CAM cell is disconnected from the ML. These transistors are used to disconnect the CAM cell from the match line (ML) during a search operation, which prevents false matches caused by the internal capacitance charging at the beginning of each search pulse. In the program mode, when these transistors are off, any input at the data line (DL) won't affect the voltage at the match line (ML), which enables the use of the ML voltage in the programming circuit to write to an individual memristor.
\end{enumerate}
\subsection{The LUT-based memristor programming algorithm}
Traditionally, the Program-Verify Algorithm is the state-of-the-art algorithm used to program memristors\cite{b11}\cite{b12}. However, while utilizing this algorithm, frequent switching between searching and programming is necessary, which leads to high programming accuracy but also results in high dynamic power consumption and lower programming speed. Besides, it requires stricter timing in peripheral circuits. Here, we present a LUT-based algorithm that can perform programming and verification simultaneously.
\begin{figure}[htbp]
\centerline{\includegraphics[scale=0.16]{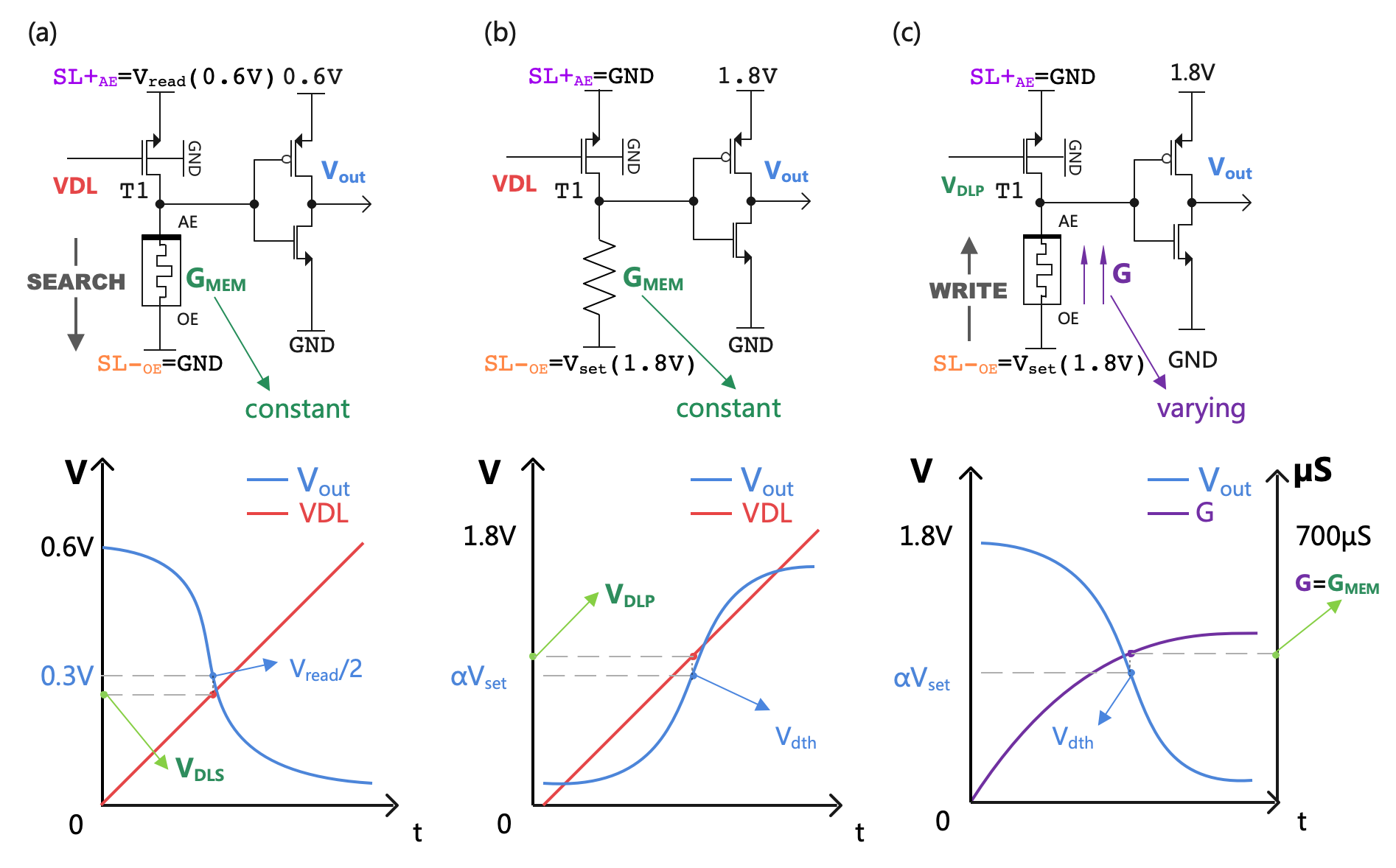}}
\caption{(a) Memristor comparator circuit in search mode with $V_{read}=$ \SI{0.6}{\volt}. If a sweeping voltage is applied at the transistor gate, the output voltage $V_{out}$ will decrease from $V_{read}$ to 0. When the intermediate voltage between the transistor and memristor equals $V_{read}/2$, the corresponding $V{DL}$ is marked as $V_{DLS}$. (b) A CMOS circuit with $V_{set}=$\SI{1.8}{\volt}. The resistor in this circuit has a fixed conductance $G_{mem}$, which is the same as the memristor conductance in circuit(a). If a sweeping voltage is applied at the transistor gate, the output voltage $V_{out}$ will increase from 0 to $V_{set}$. When $V_{out}=\alpha V_{set}$ ($0<\alpha<1$), the corresponding $V_{DL}$ is saved as $V_{DLP}$. This circuit is used to help understand the relationship between $V_{DLS}$, the boundary voltage in circuit (a) and $V_{DLP}$, the boundary voltage in circuit (b) (c) A memristive circuit with $V_{set}=$\SI{1.8}{\volt}. If a DC voltage $V_{DLP}$ is applied at the transistor gate, the output voltage $V_{out}$ will decrease from $V_{set}$ to 0 due to the increasing memristor conductance. When $V_{out}=\alpha V_{set}$ ($0<\alpha<1$), the corresponding memristor conductance in circuit (c) should be the same as the conductance of the resistor in circuit (b).}
\label{fig2}
\end{figure}
\par
The demonstration of the proposed algorithm first starts with the analysis of the circuit in search mode shown in Fig.~\ref{fig2}~(a). This circuit is called the "Memristor Comparator", which is present in the LB cell and the HB cell of the 10T2M aCAM cell. In circuit (a), the memristor is supplied with a low $VDD$ ($V_{read}=$ \SI{0.6}{\volt}), which means the conductance of the memristor remains is close to constant. To simplify the calculations, the inverter in this circuit is assumed as ideal ($V_{out}=VDD-V_{in}$). First a linear voltage sweep is applied as transient simulation ranging from $0<V_{DL}<V_{read}=$ \SI{0.6}{\volt} at the input $V_{DL}$ of the aCAM cell. By increase of $V_{DL}$ the output voltage $V_{out}$ will drop from a high voltage to a lower voltage, since the intermediate voltage between transistor and memristor will rise. If the resistor and the transistor reach the same conductance, ideally $V_{out}$ will drop from $V_{read}$ to 0. During the simulation, the $V_{DL}$ value which leads $V_{out}$ drop or rise is the desired programming value since it represents the boundary voltage to which the input is compared. For schematic (a), this $V_{DL}$ value is named $V_{DLS}$. If the transistor body effects are neglected, the drain current of the transistor at the $V_{DLS}$ point can be calculated as below:
    \begin{equation}
        \begin{split}
            I_D &=K^{\prime}\frac{W}{L}\left[\left(V_{DLS}-V_{T}\right)\frac{V_{read}}{2}-\frac{1}{2}\left(\frac{V_{read}}{2}\right)^2\right]\\
            &=\frac{V_{read}}{2}G_{mem}
        \end{split}
        \label{DLboundS}
    \end{equation}
\par
Then the relationship between $V_{DLS}$ and the memristor conductance ($G_{mem}$) can be deducted from the above equations:
\begin{equation}
    V_{DLS}=G_{mem}\frac{L}{W}\frac{1}{K^{\prime}}+V_{read}+V_T
    \label{lineardlmem}
\end{equation}
\par
Equation.~\ref{lineardlmem} demonstrates that the relation of the conductance of the memristor within the memristor comparator from Fig.~\ref{fig1} to the resulting boundary votlage $V_{DLS}$ is linear. This circumstance can be utilized to simplify programming by allowing for the programming of a boundary voltage instead of a conductance within the memristor comparator.  Conclusion \circled{\scriptsize1})
\par
Secondly, schematic (b) is compared with schematic (a) to analyze the relationship between $V_{DLP}$ and $V_{DLS}$, since in programming mode, when sweeping the $V_{DL}$ a certain specific boundary voltage $V_{DLP}$ can be observed. Circuit (b) is a CMOS circuit that substitutes the memristor in (a) with a resistor of the same conductance to assume it as a constant device. $SL-_{OE}=$ \SI{1.8}{\volt} and $SL+_{AE}=$ \SI{0}{\volt}, which means that a set voltage is applied and the transistor T1 in circuit (b) can reach saturation region. Now a simulation is performed by inserting a linear voltage sweep ($0<V_{DL}<V_{set}=$ \SI{1.8}{\volt}) at the transistor gate. With sweep of $V_{DL}$, the output voltage will rise from a low voltage to a high voltage. After this the  $V_{DL}$ is saved which makes the $V_{out}$ rise to a certain value $V_{dth}=\alpha V_{set}$ ($0<\alpha<1$), and name this $V_{DL}$ value as the boundary voltage in programming mode $V_{DLP}$. Assuming that the inverter in this circuit is ideal ($V_{out}=VDD-V_{in}$) and neglecting the transistor body effects, then the boundary voltage in programming mode can be calculated as below:
\begin{equation}
    \left(1-\alpha\right)V_{set}=\frac{1}{G_{mem}}\left[\frac{K^{\prime}}{2}\frac{W}{L}\left(V_{DLP}-V_T\right)^2\right]
    \label{DLboundP}
\end{equation}
\par
Based on Equation.~\ref{DLboundS} and Equation.~\ref{DLboundP}, the relationship between $V_{DLS}$ and $V_{DLP}$ can be written as:
\begin{equation}
    V_{DLS}=\frac{\left(V_{DLP}-V_T\right)^2}{2(V_{set}-V_{dth})}+\frac{1}{4}V_{read}+V_T
    \label{DLrealation}
\end{equation}
\par
The equation above demonstrates the relationship between $V_{DLS}$ and $V_{DLP}$. This allows a specific boundary voltage to be programmed in set mode by retrieving the related boundary voltage in read mode, $V_{DLS}$, from a lookup table.(Conclusion \circled{\scriptsize2})
\par
The schematic from Fig.~\ref{fig1}~(c) the 1T1R is again realized with a memristor. In this case, the input voltage $V_{DL}$ is not swept; instead, a constant DC voltage, $V_{DLP}$, is applied to the T1 gate. As $SL-_{OE}$ is set to $V_{set}$, the memristor comparator operates in writing mode. Now start a simulation for both circuit (b) and circuit (c). In circuit (b), $V_{DL}$ will sweep and the output voltage $V_{out}$ will start to rise as discussed before. In circuit (c), the memristor conductance will increase because the memristor will be programmed under a high $VDD$ and $V_{set}$. Thus, the continuously rising conductance will pull down the output voltage. When the output voltage of circuit (c) is decreased to $V_{dth}$, it's obvious that at this moment circuit (c) has the same working points as circuit (b) when the output voltage of (b) rises also to $V_{dth}$. Hence, when $V_{out}$ reaches $V_{dth}$ in circuit (c), the memristor in this circuit has a conductance of $G_{mem}$, which is the same as the conductance of the resistor in circuit (b).
\par
The above deductions prove that when the output voltages of these two circuits are the same, the memristor conductance in circuit (c) equals to the resistor conductance in circuit (b), and thus at this moment Equation.~\ref{DLrealation} originally calculated for circuit (b) is also valid for circuit (c) (Conclusion \circled{\scriptsize3}).
\par
Based on Conclusion \circled{\scriptsize1},\circled{\scriptsize2} and \circled{\scriptsize3}, a deduction can made be that programming the memristor conductance within circuit (a) can be switched to using a $V_{DLP}$ as input in circuit (c), and wait until the output voltage drops to $V_{dth}$. (Conclusion \circled{\scriptsize4}) It's obvious that circuit (a) and circuit (c) can represent the working status of the search mode and the program mode of aCAM cells. Consequently, Conclusion \circled{\scriptsize4} provides an algorithm to program the memristors. Using this algorithm the memristor conductance can be calculated based on Equation.~\ref{lineardlmem} and Equation.~\ref{DLrealation}:
\begin{equation}
    G_{mem}=K^{\prime}\frac{W}{L}\left[\frac{\left(V_{DLP}-V_T\right)^2}{2\left(V_{set}-V_{dth}\right)}-\frac{3}{4}V_{read}\right]
    \label{LUTequation}
\end{equation}
\par
Equation.~\ref{LUTequation} demonstrates that for the same memristor comparator circuit, a LUT between programming voltage ($V_{DLP}$) and the programmed memristor conductance ($G_{mem}$) can be built. An example LUT is shown in Table.1. In this table $V_{dth}$ is set at \SI{1.2}{\volt}. The equations above are calculated under the assumption that the transistor operates with a common long-channel quadratic function in the saturation region. In reality, the equation between $V_{DLP}$ and $G_{mem}$ depends on the transistor technology. In this paper, the simulations use TSMC 28nm technology. The transistor follows a linear function in the saturation region, and thus the programmed memristor conductance shows a linear relationship with the programming voltage.

\subsection{Programming circuit}

\begin{figure*}[h]
\centerline{\includegraphics[scale=0.18]{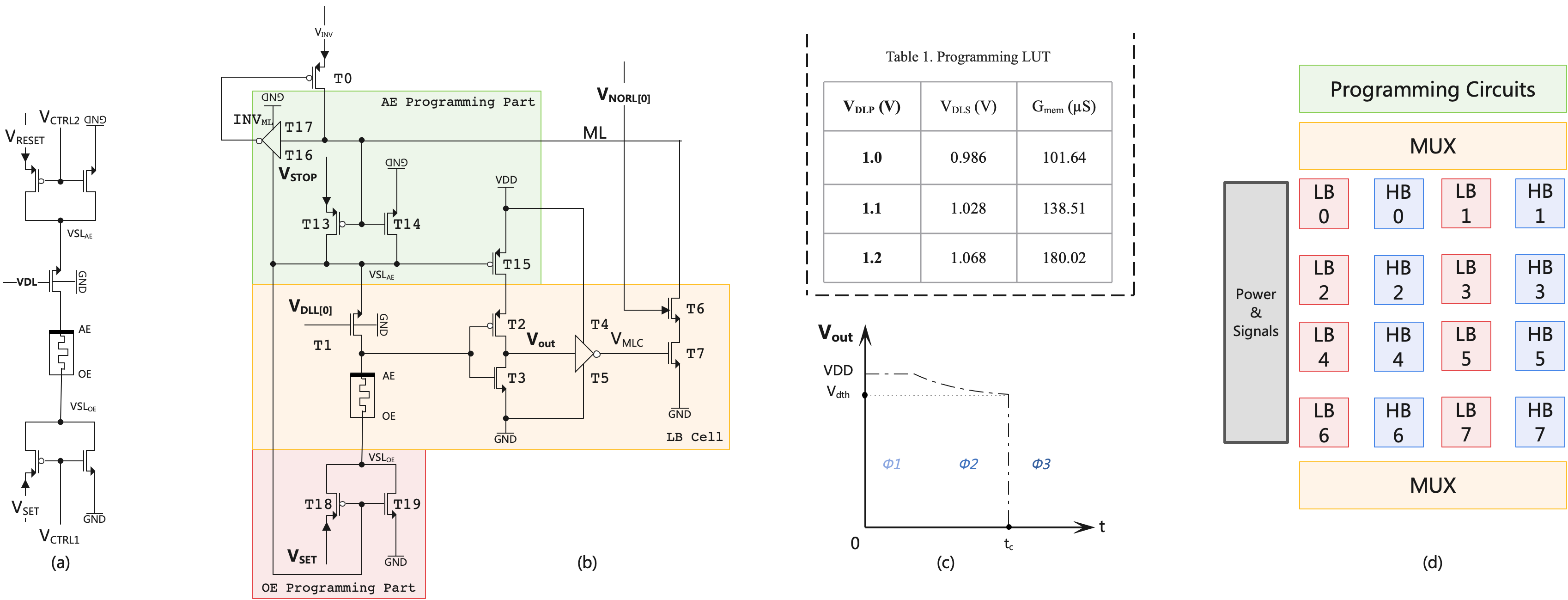}}
\caption{(a) 1T1R set-reset periphery circuit. If $V_{CTRL1}$ is high and $V_{CTRL0}$ is low, a reset of the memristor is performed (reset state); If $V_{CTRL1}$ is low and $V_{CTRL0}$ is high, a set operation will be performed (setting state). (b) LB programming circuit. This circuit receives three power signals ($V_{STOP}$, $V_{SET}$ and $V_{INV}$) and two control signals ($V_{NORL[0]}$, $V_{DL\_lb[0]}$) from an external power source, within which $V_{DL\_lb[0]}$ is the only signal that decides the conductance of the programmed memristors. The basic idea behind these circuits is to quickly switch the set-reset peripheral circuit from the setting state to the resetting state when $V_{out}$ reaches a threshold ($V_{dth}$ in Fig.~\ref{fig3}d). However, because of the higher memristor reset thresholds and the body effects of the transistor, a smaller $V_{STOP}$ used in the programming circuits won't actually reset the device but merely stop the setting process. (c) Expected $V_{out}$ transient response. The three phases of the programming circuit come from the changes of $V_{out}$. (d) Simplified aCAM array architecture. The proposed array consists of eight 10T2M aCAM cells (8 LB cells and 8 HB cells) in four rows and two columns, which means the array can store four different words at the length of two. All the aCAM cells are programmed using one single programming circuit. The array receives power and control signals from an external power source.}
\label{fig3}
\end{figure*}

To achieve the LUT-based memristor programming algorithm proposed above, the programming circuits for 10T2M LB cell is designed in Fig.~\ref{fig3}b. For all descriptions below, please refer to this diagram. Because the LB programming circuit and the HB programming circuit have similar circuit behaviors, only the programming operations in the LB cell will be demonstrated in this subsection. To set the memristor, the programming process can be roughly divided into 3 phases:
\par
In the first phase ($\phi$1:Prepare Phase), $V_{INV}$ is charged to a high voltage level (HIGH), and $V_{NORL[0]}$ must remain at a low voltage level (LOW). In this case, the ML voltage $V_{ML}$ is locked at HIGH. This will make $VSL_{AE}$ tend to connect the GND and $VSL_{OE}$ tend to connect $V_{SET}$. Now the whole circuit is ready for setting the memristor.
\par
At the beginning of $\phi$2 (Set Phase), $V_{STOP}$, $V_{NORL[0]}$ and $V_{SET}$ should be charged with corresponding high voltage. $V_{DL\_lb[0]}$ also needs to receive the programming voltage. After these signals get stable the conductance of the memristor will gradually increase, making $V_{out}$ descend slowly until reaches its threshold ($V_{dth}$), as shown in Fig.~\ref{fig3}c.
\par
After $V_{out}$ reaches its threshold $V_{dth}$, then the circuit comes to $\phi$3 (Stop Phase). When $V_{out}$ tends to be pulled down, $V_{MLC}$ on the other hand, will continuously rise to HIGH, which will pull down $V_{ML}$. In the circuit, the trend of ML pull-down will be strengthened by the ML inverter ($INV_{ML}$) and the ML PMOS (T0). At the same time, because ML controls the set-reset peripheral circuit, a lower $V_{ML}$ will lead to a higher $VSL_{AE}$ and a lower $VSL_{OE}$, which will switch the set-reset peripheral circuit tile to the resetting state. Besides, a rising $VSL_{AE}$ will reduce the source voltage of T15 and to a further extent reduce $V_{out}$. With the above positive feedback, $V_{out}$ drops to LOW at an extremely fast speed.
\par
To reset the memristor, a high voltage needs to apply at $V_{NORL[0]}$, then very high voltage needs to apply at $V_{DL\_lb[0]}$ and $V_{STOP}$. After signals are set up, the memristor can be reset to the lowest possible conductance.

\begin{figure*}[h]
\centerline{\includegraphics[scale=0.088]{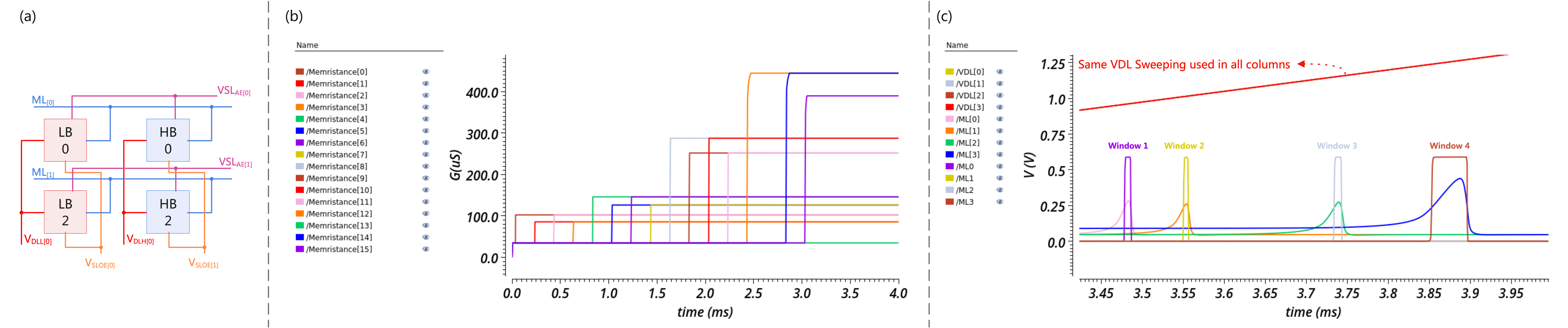}}
\caption{(a) A part of wire connections of 2 aCAM cells. Because DL (in columns) and $SL+_{AE}$ (in rows) need to be activated in the Reseting Mode, other cells won’t be affected if an LB or HB cell is chosen to reset. Similarly, in the Writing Mode, ML (in rows) and DL (in columns) need to be used, thus setting one single LB or HB cell won’t set other nearby cells. (b) The proposed aCAM array is programmed with 4 different words. This diagram shows the transient response of the memristor conductance within 8 LB cells and 8 HB cells. (c) When the same $V_{DL}$ sweeping voltage for searching is applied, the array programmed as Fig.~\ref{fig4}b shows four distinct analog windows (matching ranges). }
\label{fig4}
\end{figure*}

\begin{figure*}[htbp]
\centerline{\includegraphics[scale=0.11]{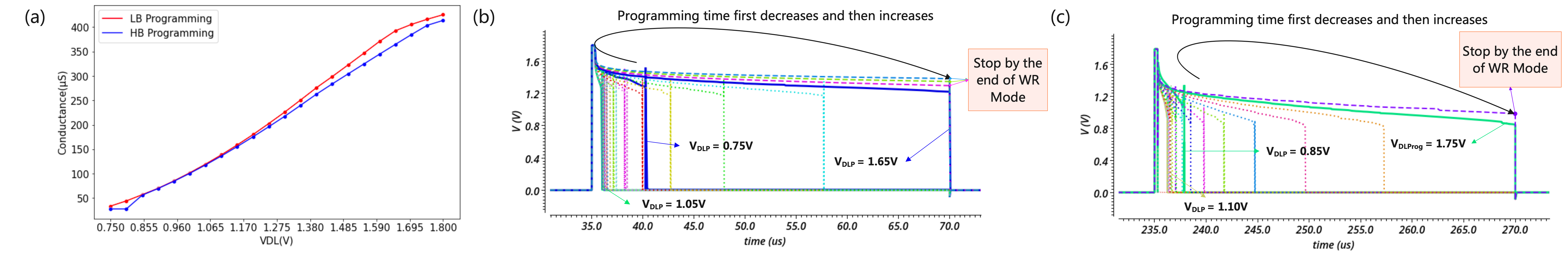}}
\caption{(a) One single LB and HB Cell programming results. In the simulations of single-cell programming, a DC $V_{DL}$ ranging from \SI{0.75}{\volt} to \SI{1.8}{\volt} (with a fixed step at \SI{0.05}{\volt}) is used to set the memristors. The maximum programming period is 35$\mu$s. It's obvious that the programmed conductance shows a linear relationship within a certain range of programming $V_{DL}$. (b) LB cell $V_{out}$ transient response. The programming time of the memristor first decreases and then increases with the rising $V_{DL}$ (c) HB cell $V_{out}$ transient response. The programming time of the memristor also first decreases and then increases with the rising $V_{DL}$}
\label{fig5}
\end{figure*}

\subsection{aCAM array design}
For convenience, if the aCAM cell (Fig.~\ref{fig2}) is drawn with LB and HB blocks, a simplified architecture of the proposed 8-Cell aCAM array design is shown in Fig.~\ref{fig3}d. The array has four working modes: Writing Mode (WR), Resetting Mode (RST), Sweeping Mode (SW), and Verify Mode (VR). Besides, the array has 2 states: the programming state and the searching state. In the programming state, the array is programming the memristor in LB or HB cells to different conductance levels, followed by some checking and verification work. In the searching state, the parallel $V_{DL}$ is input to the array and the array will check if there is a match in a row.
\par
The functions of each working mode are listed below:
\begin{enumerate}
\item In the WR, the conductance of the memristor in an LB cell or an HB cell is set to the desired value.
\item In the RST, the memristor is reset to the lowest possible conductance.
\item In the SW, a voltage sweep between $0$ to \SI{0.6}{\volt} is used at $V_{DL}$. This mode is used to check whether the aCAM cell works normally.
\item In the VR, \SI{1.9}{\volt} is applied at $V_{DL}$. This mode is used to precisely measure the memristor conductance.
\end{enumerate}
\par

\section{Simulation Results}
An attractive feature of the proposed analog feedback-controlled programming circuits is the possibility to program the memristor conductance only using the data line voltage $VDL$ as a input parameter. To test whether the proposed memristor programming circuits and the aCAM array can achieve the programming functions, simulations for both single-cell programming and array programming have been performed. In the single-cell programming, the results Fig.~\ref{fig5}a shows that the programmed conductance shows a linear relationship within a certain range of programmable $V_{DL}$, which also corresponds to the memristor comparator's dynamic range. Fig.~\ref{fig5}b and Fig.~\ref{fig5}c show that the programming process under high programming $V_{DL}$ will be stopped earlier by the end of the WR Mode.
\par
Fig.~\ref{fig4}c shows the memristor conductance changes of all 8 aCAM cells. Because before setting an LB or HB cell the memristor in this cell is reset to the lowest conductance, the figure proves that setting or resetting a specific cell won't affect other cells. Figure.~\ref{fig4}d illustrates the function of aCAM. In the Programming State, $8$ aCAM cells in $4$ rows are programmed with $4$ different matching ranges (windows). Then in the Searching State, when a sweeping $V_{DL}$ is used as input, it's clear to see the shapes of these windows.
\section{Conclusion}
In this article, we implement an analog feedback-controlled memristor programming circuit, which leverages the natural relationship between memristor conductance and the programming data line boundary voltage in memristor CAMs. The simulation results show that the proposed programming circuits meet the expected transient response and could achieve the LUT-based memristor programming algorithm. In addition, we manage to integrate eight 10T2M aCAM cells with a single programming circuit. The simulations prove that the proposed array architecture can achieve the functions of an aCAM array.
\par
Despite the benefits of the proposed hardware architectures, the memristor devices still have many non-ideal characteristics, such as device to device and cycle to cycle variability, read noise and conductance drifts, which are not considered in this article. In addition, the ML parasitics of the proposed array are not fully counted in the simulations, which may affect the final results of the read/write speed and memory density. The aforementioned non-idealities will be considered and analyzed in our future work.

%%%%%%%%%%%% Supplementary Methods %%%%%%%%%%%%
%\footnotesize
%\section*{Methods}

%%%%%%%%%%%%% Acknowledgements %%%%%%%%%%%%%
%\footnotesize
%\section*{Acknowledgements}

%%%%%%%%%%%%%%   Bibliography   %%%%%%%%%%%%%%
\clearpage
\normalsize
\bibliography{references}

%%%%%%%%%%%%  Supplementary Figures  %%%%%%%%%%%%
%\clearpage

%%%%%%%%%%%%%%%%   End   %%%%%%%%%%%%%%%%
%\end{multicols}  % Method B for two-column formatting (doesn't play well with line numbers), comment out if using method A
\end{document}